Revtex 3.0 , 8 figures uuencoded compressed tar-file
\documentstyle[preprint,aps]{revtex}
\begin{document} \draft
\title{Coherent and sequential photoassisted tunneling
through a semiconductor double barrier structure.}
\author{J. I\~{n}arrea, G. Platero}
\address{Instituto de Ciencia de Materiales (CSIC) and  Departamento de 
Fisica de la Materia Condensada C-III, Universidad Autonoma,
Cantoblanco, 28049 Madrid, Spain.}
\author{C. Tejedor}
\address{Departamento de Fisica de la Materia Condensada,
Universidad Autonoma de Madrid, Cantoblanco, 28049 Madrid, Spain.}
\maketitle   
\begin{abstract}
We have studied the problem of coherent and sequential tunneling 
through a 
double barrier structure, assisted by light considered to be present
All over the structure, i,e emitter, well and collector as in the
experimental evidence.  By means of a
canonical transformation and in the framework of the time dependent
perturbation theory, we have calculated the transmission coefficient
and the electronic resonant current.
Our calculations have been compared with experimental results 
turning out
to be in good agreement.
Also the effect on the coherent tunneling of a magnetic field
parallel to the current in the presence of light, has been considered.
\end{abstract}
\newpage
\narrowtext
\section{INTRODUCTION}
\par
Resonant tunneling\cite{1} through double barrier structures, (DBS)
has been one of the most active fields in research
 in solid state physics, both from   
theoretical and experimental standpoints. The main reason is 
that resonant
tunneling has been considered to have a great potential applicability in
electronic devices. In the same way, the interaction of an external 
time-dependent potential with resonant structures is considered to have 
very interesting applications, for instance the use of DBS
as detectors and generators of microwave radiation. In this
paper we are going to study the effect of a photon field on both coherent
and sequential tunneling current through a DBS.\\
\par
The work of Sollner et al\cite{2}, is the experimental starting 
point for 
studies on the effect of time-dependent potentials in resonant tunneling
through semiconductor microstructures: they studied the influence of 
electromagnetic
radiation on resonant tunneling current. Recently Chitta et al\cite{3}
have studied the far infrared response of double barrier 
resonant tunneling
structures. Theoretical work on tunneling devices
under the influence of a time-dependent potential has a long history. 
Tien and
Gordon\cite{4}, studied the effect that microwave radiation has on 
superconducting tunneling devices. Several authors\cite{5,6,7,8,9,10}
have investigated the effect that external AC potentials have in 
different problems.
Jonson\cite{11}, Apell et al\cite{12} and Johansson et al\cite{13} have 
studied the sequential contribution to the
tunneling through a DBS under an electromagnetic field
applied, using models based in the Transfer Hamiltonian 
formalism\cite{14}.
In all those models above, the coupling between electrons 
and the electromagnetic
field is considered to take place just in a part of the structure: 
in most of them in the well, and in the case of Apell et al\cite{12} 
 in the emitter and collector, but in none of them affecting 
 the whole structure.\\
\par
In this paper we have calculated how the transmission 
coefficient and the current for electrons in a DBS are changed
due to the presence of light in the whole structure. In order 
to do that 
we have developed a quantum mechanical formalism to find the expression
for the electronic state dressed by photons and we have calculated
the resonant tunneling current under the influence of an external
electromagnetic field. This quantum mechanical formalism based in
a canonical transformation and in the time dependent perturbation 
theory,
has been aplied to coherent and sequential tunneling processes, 
and the
results we have obtained are in good agreement with the available
experiments\cite{3}. The case of coherent resonant tunneling assisted 
by light in the presence of a magnetic field parallel to the current, 
has been also studied.\\
\par
This paper is organized as follows: In Sec II, we discuss and 
develop the theoretical formalism. In Sec III.a and b, we applied that 
formalism to coherent and  sequential tunneling respectively. 
In Sec IV, our results for both types of tunneling 
for different frequencies, external electromagnetic fields and
magnetic fields are presented and compared with 
experimental\cite{3} results. We summarize our conclusions in Sec V.\\

\section{ELECTRONIC STRUCTURE OF THE SYSTEM WITH LIGHT.}
\par
The quantum mechanical Hamiltonian for an electron in the presence 
of an electromagnetic field represented by a plane electromagnetic 
wave of wave vector $\vec{k}$, parallel to the $x$ direction and polarized 
in the $z$ direction
$\vec{E}=(0,0,F)$, (see fig 1), can be written as:
\begin{eqnarray}
H_{tot}=(1/2m^{*})(\vec{P}+e\vec{A}(\vec{R},t))^{2}+V(\vec{R})+\hbar w a^{+}a
\end{eqnarray} 
In our problem we apply an external bias, such that the electrostatic
and barriers potential depends only on the $z$ 
direction so we take the potential $V(\vec{R})$ as $V(z)$. 
In the Coulomb
gauge $\vec{\bigtriangledown} . \vec{A}    =0 $ then (1) becomes:
\begin{eqnarray}
H_{tot} =  P^{2}/2m^{*} + (e/m^{*}) \vec{P}.\vec{A}(\vec{R},t)+
(e^{2}/2m^{*})A^{2} (\vec{R},t)+V(z)+\hbar w a^{+}a
\end{eqnarray}
In our case the vector potential operator 
$\vec{A}(\vec{R},t)=A_{z}(x,t)$. 
 In general, $A^{2}(R,t)$ is negligible compared to the 
$(e/m^{*})\vec{P}.\vec{A}(\vec{R},t)$ term, therefore we can write in second
quantization for the total Hamiltonian:
\begin{eqnarray}
H_{tot}=H_{e}^{0}+H_{ph}^{0}+W_{D}(t)+W_{OD}(t)
\end{eqnarray}
where
\begin{eqnarray}
H_{e}^{0}&=&\sum_{k} \epsilon_{k} c_{k}^{+} c_{k} \\
H_{ph}^{0}&=&\hbar w a^{+}a\\
W_{D}(t)&=&\sum_{k} [(e/m^{*})<k|P_{z}|k> c_{k}^{+} c_{k}
  (\hbar/2\epsilon Vw)^{1/2} (a e^{-iwt} +a^{+} e^{iwt})]\\
W_{OD}(t)&=& \sum_{k}\sum_{k^{'}\neq k} [(e/m^{*})<k^{'}|P_{z}|k>
  c_{k^{'}}^{+} c_{k} (\hbar/2\epsilon Vw)^{1/2}
   (a e^{-iwt} + a^{+} e^{iwt})]
\end{eqnarray}
where $A_{z}(x,t)=(\hbar/2\epsilon Vw)^{1/2} 
\vec{\varepsilon_{z}} (a e^{-iwt} + a^{+} e^{iwt}) $ being $w$ the photon
frequency, the wave vector of the electromagnetic
field has been neglected, and where the term $e^{-iEt}$ is already
included in the state vector $|k>$ . $H_{e}^{0}$ is the independent, 
electronic Hamiltonian and includes the double barrier
potential and the external applied bias, therefore 
the eigenstates of $H_{e}^{0}$, $\Psi_{0}(k)$, are the tunneling states 
for bare electrons.
$H_{ph}^{0}$, is the photon field Hamiltonian whithout coupling with
electrons and $W_{D}$ and $W_{OD}$, describe the coupling 
between electrons
and photons in the total Hamiltonian. 
We separate the coupling term in the
"diagonal" (6) and the "off-diagonal" (7) contributions because
 we are going to be interested in problems where a quasi-localized state
is connected by the electromagnetic field with a continuum of extended
states. Therefore $W_{OD}$, can be treated in first order time dependent
perturbation theory. For problems in which two o more quasi-localized
states should be connected by the light, the method could not be applied
in the same way, requiring some generalization.  
. Therefore the total Hamiltonian can be written as:
\begin{eqnarray}
H_{tot}=H_{D}(t)+W_{OD}(t)
\end{eqnarray}
where $H_{D}(t)=H_{e}^{0}+H_{ph}^{0}+W_{D}(t) $
\\
\par
The hamiltonian $H_{D}$, can be solved exactly
Considering a canonical transformation\cite{11},\cite{15}.  
It allows to obtain the exact electronic wave function dressed by photons: 
$\Psi_{D}(k)=U^{+}\Psi_{0}(k)$,  where $\Psi_{0}(k)$ is 
the electronic double
barrier eigenstate with no photon field present in the sample. 
Once we have obtained the eigenstates for $H_{D}$, we apply the time dependent
perturbation theory in order to treat the $W_{OD}$ 
term. The operator $U$ for the canonical transformation is given by 
$U=e^{s}$, and $s$ can be written as :
\begin{eqnarray}
s&=&\frac{e}{m^{*}\hbar w}(\frac{\hbar}{2\epsilon Vw})^{1/2}
 <P_{z}>c_{k}^{+}c_{k} (a^{+} e^{iwt} - a e^{-iwt})\nonumber\\
 & & \nonumber\\
 &=&\frac{M}{\hbar w} c_{k}^{+}c_{k} (a^{+} e^{iwt} - a e^{-iwt})
\end{eqnarray}
The Hamiltonian under this transformation becomes:
\begin{eqnarray}
\tilde{H}_{D}&=&\tilde{c}_{k}^{+}\tilde{c}_{k} (\epsilon_{k}-\frac{M^{2}}{
\hbar w})+ \hbar w \tilde{a}^{+} \tilde{a} 
\end{eqnarray}
where $\tilde{a}^{+}=a^{+}-\frac{M}{\hbar w} c_{k}^{+}c_{k}$
and  $\tilde{a}=a-\frac{M}{\hbar w} c_{k}^{+}c_{k}$ \cite{15}.
In the transformed Hamiltonian $\tilde{H}_{D}$ the electrons and photons are
not coupled any more and the electronic eigenvalues are shifted in
$\Delta = \frac{M^{2}}{\hbar w}$ which is negligible 
with respect to the free electron eigenvalues. 
Finally we can write the exact eigenstate for $H_{D}$ in terms
of the electric field intensity $F$ \cite{16}:
\begin{eqnarray}
\Psi_{D}(k)=exp{[\frac{-ieF}{m^{*}\hbar w^{2}}<P_{z}>sin(wt)]}\Psi_{0}(k)=
  \Psi_{0}(k) \sum_{n=-\infty}^{\infty}J_{n}(\beta_{k}) e^{-inwt}
\end{eqnarray}
where $\beta_{k}=\frac{eF<P_{z}>}{m^{*}\hbar w^{2}}$. 
\\
\par                                                    
At this point, and in order to obtain the total wave function where 
non-diagonal terms $(k^{'} \neq k)$ are included, 
we consider time dependent perturbation theory up to first order. 
For that purpose we calculate the total wave
function time dependent coefficients, which are given by:
\begin{eqnarray}
C_{f}^{(1)}(t)=\lim_{\alpha \rightarrow 0} \int_{-\infty}^{t}
  (1/i \hbar) <\Psi_{D}(k^{'})|W(k)|\Psi_{D}(k)> e^{\alpha t} dt
\end{eqnarray}
where
\begin{eqnarray}
W(k)=(eF/m^{*} w) \sum_{k^{'}}<k^{'}|P_{z}|k>
c_{k^{'}}^{+} c_{k} cos(wt)
\end{eqnarray}
Since we consider first order time dependent perturbation theory, 
we keep only the $J_{0}$ Bessel
functions terms because if we took the $J_{1}$ terms or terms of higher
order in the Bessel functions , 
that would mean to consider second or higher order processes 
giving a very small contribution to the total wave function. 
 Due to that we will see below that only one photon absorption and
emission processes 
are considered in our formalism. From (11), (12) and (13) we have :
\begin{eqnarray}
C_{f}^{(1)}(t)&=&\frac{-eFL}{4\pi \hbar^{2}w} 
\int_{0, k^{'}\neq k}^ {\infty} J_{0}(\beta_{k^{'}})J_{0}(\beta_{k})
<k^{'}|P_{z}|k>/k^{'} \times \nonumber\\
	      & &[e^{i(w_{k^{'}k}+w)t}(PP\frac{1}{w_{k^{'}k}+w}+
			i\pi\delta (w_{k^{'}k}+w)+ \nonumber\\
	      & &e^{i(w_{k^{'}k}-w)t}(PP\frac{1}{w_{k^{'}k}-w}+
			i\pi\delta (w_{k^{'}k}-w)] dw_{k^{'}}
\end{eqnarray}
In the calculation of this integral, the principal part term 
results to be
negligible compared to the $\delta$ term. If we carry out that integral,
taking the above into account we can obtain for the coefficients:
\begin{eqnarray}
C_{1,(-1)}^{(1)}=(-ieFL/4 \hbar^{2}w)J_{0}(\beta_{k_{1,(-1)}})J_{0}
(\beta_{k_{0}}) <k_{1,(-1)}|P_{z}|k_{0}>/k_{1,(-1)}
\end{eqnarray}
\\
Therefore, denoting by $k_{0}$, the wave vector of the initial electron we can
write for the total wave function:
\begin{eqnarray}
\Psi(t)=\alpha[\Psi_{D}(k_{0})+C_{1}^{(1)}(t)\Psi_{D}(k_{1})+C_{-1}^{(1)}(t)
   \Psi_{D}(k_{-1})]
\end{eqnarray}
The normalization constant 
$\alpha=1/\sqrt{1+|C_{1}^{(1)}(t)|^{2}+
|C_{-1}^{(1)}(t)|^{2}}$, 
 in eq. (16) guarantees current conservation, in other words, in
the presence of a barrier the addition of transmision and reflection 
probabilities will give 1, as discussed below.
$\Psi_{D}(k_{0})$ is the "dressed" 
reference state and
 $\Psi_{D}(k_{1})$, and $\Psi_{D}(k_{-1})$, are the two 
coupled "dressed"
states due to one photon absorption and emission processes and
$C_{1,(-1)}^{(1)}(t)$, are the corresponding coefficients for $\Psi(t)$ 
coming from the 1 photon absorption (emission) processes, 
$(w_{1,-1}=w_{k_{0}}\pm w)$.
\\
\section{LIGHT ASSISTED TUNNELING THROUGH A DBS}
\subsection{Coherent tunneling.}
\par
Now we use the above electronic structure to analyze the problem of 
coherent resonant tunneling in a DBS assisted by light. Before
turning on the light we are going to calculate, applying the transfer
matrix technique, the transmission coefficient for a double barrier.
First of all we write in this framework,
the wave function in the emitter and collector regions for an electron 
($\Psi_{e0}(k_{e})$ and $\Psi_{c0}(k_{c})$, respectively)
crossing the double barrier (see fig 1):
\begin{eqnarray}
\Psi_{e0}=1/\sqrt{L}(e^{ik_{e}z}+re^{-ik_{e}z}) e^{ik_{x}x} 
e^{ik_{y}y} e^{-iw_{0}t} \\
\Psi_{c0}=t/\sqrt{L}e^{ik_{c}z} e^{ik_{x}x} e^{ik_{y}y} e^{-iw_{0}t}
\end{eqnarray}
where $k_{e}$, and $k_{c}$, are the electronic wave vector 
perpendicular components in the emitter and collector 
respectively. 
 The incident and transmitted currents are $J_{i}=\frac{\hbar 
k_{e}}{m^{*}\sqrt{L}}$, and
$J_{t}=\frac{\hbar k_{c}}{m^{*}\sqrt{L}}|t|^{2}$, respectively so 
that the transmission coefficient is $T_{0}=\frac{k_{c}}{k_{e}} |t|^{2}$.,
where the factor $|t|^{2}$ is calculated by means of the
Transfer matrix formalism, i.e. imposing the boundary continuity of wave
function and current at the barrier interfaces\cite{17}.\\
\par
If now we turn on the light, our state is transformed in the electron- 
photon wave function $\Psi(t)$ (16). From that  
we calculate the new incident and transmitted currents, and after some
algebra the transmission coefficient in the presence of light becomes:
\begin{eqnarray}
T&=&T_{0}/(1+k_{1}/k_{0}|C_{1}^{(1)}|^{2}+k_{-1}/k_{0}
|C_{-1}^{(1)}|^{2})+\nonumber\\
 & &T_{1}|C_{1}^{(1)}|^{2}/(k_{0}/k_{1}+|C_{1}^{(1)}|^{2}+k_{-1}/k_{1}
 |C_{-1}^{(1)}|^{2})+\nonumber\\
 & &T_{-1}|C_{-1}^{(1)}|^{2}/(k_{0}/k_{-1}+k_{1}/k_{-1}
|C_{1}^{(1)}|^{2}+ |C_{-1}^{(1)}|^{2})
\end{eqnarray}
A similar expresion can be obtained for the reflexion coefficient : 
\begin{eqnarray}
R&=&R_{0}/(1+k_{1}/k_{0}|C_{1}^{(1)}|^{2}+k_{-1}/k_{0}
|C_{-1}^{(1)}|^{2})+\nonumber\\
 & &R_{1}|C_{1}^{(1)}|^{2}/(k_{0}/k_{1}+|C_{1}^{(1)}|^{2}+k_{-1}/k_{1}
 |C_{-1}^{(1)}|^{2})+\nonumber\\
 & &R_{-1}|C_{-1}^{(1)}|^{2}/(k_{0}/k_{-1}+k_{1}/k_{-1}
|C_{1}^{(1)}|^{2}+ |C_{-1}^{(1)}|^{2})
\end{eqnarray}
where $R_{0}$, $R_{1}$ and $R_{-1}$,
($T_{0}$, $T_{1}$ and $T_{-1}$), 
 are  the standard coherent  double barrier
reflexion (transmission )  coefficients, 
evaluated at the reference energy, at one
photon above and at one photon below the reference energy, respectively.
 This expression for the reflexion coeffient verifies  the  current
conservation:  $|T|^{2}+|R|^{2}=1$, it means that the
probability for an electron to tunnel with no photon  absorption or emission 
is smaller than the corresponding with no light present in the sample.
It is due to the finite probability associated to emission and
absorption processes and  it is  a consequence of the unitarity \cite{18}, 
and comes directly from the normalization of the total electronic wave
function where one photon absorption and emission processes are
considered.  
 As the electromagnetic field intensity increases, the inelastic
processes are more probable and therefore, the elastic or direct
tunneling has a smaller probability than for low field intensities . 
In order to analyze the dc current, which is the only one observed 
in the experiments\cite{3}, we have made a time average so that
no interference terms appear. Finally the total electronic 
current can be written as:
\begin{eqnarray}
J_{tot}=\frac{em^{*}}{2\pi^{2}\hbar^{3}} \int (f(E)-f(E+V_{f}))TdE_{z}
\int dE_{p}
\end{eqnarray}
being $f$ the Fermi function, $E_{p}$ the electronic energy parallel part 
and $V_{f}$ the external DC applied bias. 
\\
\par
We can now consider the problem of adding a magnetic field
$\vec{B}$, parallel to the current direction, i.e., the $z$
direction. In the Landau gauge : $\vec{A_{B}}=(-yB,0,0)$.
The effect of this magnetic field is to change the 
parallel part of the density of states and due to that instead
of a continuum of states we have now the Landau levels
ladder. The Hamiltonian for an electron in the presence of an
electromagnetic field in the configuration considered above and
a magnetic field parallel to the current can be describe in 
second quantization as:
\begin{eqnarray}
H_{tot}=H_{e}^{0}+H_{ph}^{0}+W_{D}(t)+W_{OD}(t)
\end{eqnarray}
$H_{ph}^{0}$, $W_{D}(t)$,and $W_{OD}(t)$, are exactly the same
as the ones described in the general formalism, but 
 $H_{e}^{0}$  has been transformed due to the presence of the
magnetic field and can be written now in second quantization as:
\begin{eqnarray}
H_{e}^{0}=\sum_{k} \epsilon_{z} c_{z}^{+} c_{z}+
     \hbar w_{c}(a_{B}^{+} a_{B} + 1/2)
\end{eqnarray}
where $B$  is the magnetic field intensity,
$w_{c}$  is the cyclotron frequency: $w_{c}=eB/m^{*}$, 
$a_{B}^{+}$, and $a_{B}$ are the creation and destruction    
operators for the Landau states, and $\epsilon_{z}$ is 
the electronic energy perpendicular part. 
With no magnetic field present in the sample the parallel component for
the electronic wave vector is conserved during the photoassisted 
tunneling process. Now as the magnetic field is switched on, 
is the Landau level index
what is conserved. The alignment of the Landau levels in
the emitter and in the well with the same index, gives a jump
in the electronic current, giving\cite{19} eventually a sawtooth profile
for the I-V characteristic depending on the magnetic field
intensity. 
The expression for the current can be written then as:
\begin{eqnarray}
J=(2/2\pi^{2})(e/\hbar)^{2} B \sum_{n=0}^{N}\int_{(n+1/2)\hbar w}
 ^{E_{F}} dE [f(E)-f(E+V_{f})] T(E,n)
\end{eqnarray}
being $n$ the Landau level index, $N$, the maximum ocupied Landau
level index, and $T(E,n)$ the transmission 
coefficient when the photon field is present in the sample (19).
\subsection{Sequential tunneling} 
In order to study the sequential tunneling, we have developed a model
that calculates separately the current for the first and the 
second barriers, $J_{1}$, and $J_{2}$. These currents are
related to the Fermi level in the well $E_{w}$ or in 
other words, to the amount of electronic charge stored into
the well. In this model we adjust selfconsistently the Fermi
level till the currents through the first and the second barriers
result to be equal. The values calculated in this way for 
the current and the Fermi level in the well, are indeed the actual current
which is crossing the whole double barrier sequentialy and the Fermi level
corresponding to the actual amount of charge  stored
into the well. This model takes into account macroscopically the
possible 
scattering processes within the well.\\
\par 
In order to calculate $J_{1}$, and $J_{2}$ without light, we use the
transfer Hamiltonian method \cite{14}. We calculate for
the first barrier, the probability $P_{1}$ for the electron 
to cross from the emitter to the well :
\begin{eqnarray}
P_{1}=(2\pi/\hbar)(2\pi/L^{2})^{2}[\frac{\hbar^{4}k_{e}k_{w}}
{2m^{*2}L(w_{2}+(1/\alpha_{b})+
(1/\alpha_{d}))}T_{s}\delta (k_{p}^{w}-k_{p}^{e})
\delta(E_{z}-E_{tn})]
\end{eqnarray}
where $T_{s}$ is the transmission coefficient for a single barrier;
$k_{e}$ ($k_{p}^{e}$) and $k_{w}$ ($k_{p}^{w}$) are the perpendicular
(parallel) component for the electronic
wave vector in the emitter and well respectively; $E_{tn}$, is the well 
state energy referred to conduction band
bottom: $E_{tn}=E_{R}-V_{f}(w_{1}+w_{2}/2)/w_{t}$ (where $E_{R}$,
is the well state energy referred to well bottom) ; $\alpha_{b}
=\sqrt{\frac{2m^{*}(V_{0}-E_{R}+V_{f}(w_{1}+w_{2})/
2w_{t})}{\hbar^{2}}}$, $\alpha_{d}=\sqrt{\frac{2m^{*}(V_{0}-E_{R}+
V_{f}(w_{2}+w_{3})/2w_{t})}{\hbar^{2}}}$; $w_{1}$, $w_{2}$ and 
$w_{3}$  are the first barrier, well and second barrier widths and 
$w_{t}$ is the total width for the whole structure. 
It is important to stress the presence of the $\delta(E_{z}-E_{tn})$ term
in the $P_{1}$ expression. It implies that only for those emitter states which
resonate with the well state, will be possible to cross the emitter
barrier to the well and
therefore contribute to the current. With this
probability $P_{1}$, we can calculate the current $J_{1}$ that, after 
integrating in the energy is given by:
\begin{eqnarray} 
J_{1}=(e/2\pi\hbar)\frac{k_{w}T_{s}}{w_{2}+(1/\alpha_{b})+(1/\alpha_{d})}
(E_{F}-E_{tn}-E_{w})
\end{eqnarray}
where $E_{F}$ and $E_{w}$, are the Fermi level energies in the emitter and
in the well respectively. For the second barrier, we apply exactly the same
formalism and we obtain for the probability of crossing from the well
 to the collector $P_{2}$  and for the current through the second
barrier $J_{2}$:
\begin{eqnarray} 
P_{2}&=&(2\pi/\hbar)(2\pi/L^{2})^{2}[\frac{\hbar^{4}k_{w}k_{c}}
{2m^{*2}L(w_{2}+(1/\alpha_{b})+ (1/\alpha_{d}))}T_{s}
\delta (k_{p}^{c}-k_{p}^{w}) \delta(E_{z}-E_{tn})]\\
J_{2}&=&(e/2\pi\hbar)\frac{k_{w}T_{s}}{w_{2}+(1/\alpha_{b})+
(1/\alpha_{d})} E_{w}
\end{eqnarray}
where, $k_{w}$ is the perpendicular component for the
electronic wave vector in the well.\\
\par
Repeating the arguments we have made to study
the effect of the light on coherent tunneling, it is 
straightforward to extend that formalism to the sequential
tunneling case. 
Before switching on the light the electrons have just one way to get 
into the well state
from the emitter: from an emitter state which is resonant with the well state
i.e., having the $\delta(E_{z}-E_{tn})$ term in the integral. Now
when we switch on the light, the electrons have three different ways 
to tunnel through the emitter barrier to the well. The 
first one is a direct way and it corresponds to an emitter state which 
resonates with the well state, the transmission takes place whithout light
absorption or emission. The second one is through and absorption process
from an emitter state which is found at one photon energy below the resonant
well state. And finally the third way is through an emission process from
an emitter state which is found at one photon energy above the resonant
well state. For those reasons above we will have in the $J_{1}$ expression
the sum of three terms  in each one appearing a different $\delta$ function. 
The direct term
has in its expression  a $\delta[E_{z}-E_{tn}]$, the absorption term  a
$\delta[E_{z}-(E_{tn}-\hbar w)]$, and finally  the emission term a
$\delta[E_{z}-(E_{tn}+\hbar w)]$. So the final expression we have for the current
$J_{1}$:
\begin{eqnarray}
J_{1}&=&(e/2\pi\hbar)\int_{0}^{E_{F}} dE_{z} \frac{k_{w}}
{w_{2}+(1/\alpha_{b})+(1/\alpha_{d})}\times \nonumber\\
     & &[\delta[E_{z}-E_{tn}]
     \frac{T_{s,0}}{1+k_{1}/k_{0}|C_{1,0}^{(1)}|^{2}+k_{-1}/k_{0}
       |C_{-1,0}^{(1)}|^{2}}+\nonumber\\
     & &\delta[E_{z}-(E_{tn}-\hbar w)]
     \frac{T_{s,1}|C_{1,0}^{(1)}|^{2}}{k_{0}/k_{1}+|C_{1,0}^{(1)}|^{2}+
       k_{-1}/k_{1}|C_{-1,0}^{(1)}|^{2}}+\nonumber\\
     & &\delta[E_{z}-(E_{tn}+\hbar w)]
     \frac{T_{s,-1}|C_{-1,0}^{(1)}|^{2}}{k_{0}/k_{-1}+k_{1}/k_{-1}
      |C_{1,0}^{(1)}|^{2}+ |C_{-1,0}^{(1)}|^{2}}
     ] \int_{E_{w}}^{E_{F}-E_{z}}dE_{p}
\end{eqnarray}
If we make this integral, we can finally obtain for $J_{1}$:
\begin{eqnarray}
J_{1}=(e/2\pi\hbar)[\frac{k_{w,0}T_{s,0}(E_{F}-E_{tn}-E_{w})}
{w_{2}+(1/\alpha_{b,0})+(1/\alpha_{d,0})} \frac{1}{1+k_{1}/k_{0}
|C_{1,0}^{(1)}|^{2}+k_{-1}/k_{0}|C_{-1,0}^{(1)}|^{2}}+\nonumber\\
\frac{k_{w,-1}T_{s,0}(E_{F}-(E_{tn}-\hbar w)-E_{w})}
{w_{2}+(1/\alpha_{b,-1})+(1/\alpha_{d,-1})}\frac{|C_{1,-1}^{(1)}|^{2}}
{k_{-1}/k_{0}+|C_{1,-1}^{(1)}|^{2}+k_{-2}/k_{0}|C_{-1,-1}^{(1)}|^{2}}+
 \nonumber\\
\frac{k_{w,1}T_{s,0}(E_{F}-(E_{tn}+\hbar w)-E_{w})}
{w_{2}+(1/\alpha_{b,1})+(1/\alpha_{d,1})}\frac{|C_{-1,1}^{(1)}|^{2}}
{k_{1}/k_{0}+|C_{1,1}^{(1)}|^{2}k_{2}/k_{0}+|C_{-1,1}^{(1)}|^{2}}]
\end{eqnarray}
where as above, the subscript "0" means the reference  state energy that
in our case is the resonant well state energy. The subscript "1" and "-1"
mean one photon energy above and below respectively, etc.\\
\par

For the second barrier we do not have the constraint of
crossing to a specific discrete state, but what we have
now is a continuum of states in the collector. The expression
we have for the current through the second barrier in the 
presence of light is formally equal to the $J_{1}$ expression
 i.e., it is formed for the sum of three contributions, each one at
different energy.
At this point we apply the same procedure as in the case where there is
no light present, i.e., we calculate selfconsistently the Fermi level
in the well till both currents for the first and second barriers, 
result to be equal. The values obtained in this way are the 
actual photoassisted sequential current and Fermi level in the well.
\section{RESULTS}
We have performed a calculation for a GaAs-GaAlAs DBS with a well 
and barriers thicknesses of $50 \AA$ , in order to analyze the experimental
information\cite{3}. 
The electromagnetic field is polarized along the sample growth direction
(fig. 1) , 
and the carrier density $n=10^{18}$ $cm^{-3}$.
First of all we have calculated the 
total transmission coefficient for coherent tunneling , 
for a field intensity $F=4.10^{5} V/m$ and energy $\hbar w=13.6 meV$
 and for different external bias ($V_{f}=0.0 V, 0.1 V$ and 0.14 V, 
figures 2a, b, c).
The main features observed in the transmission coefficient as a function
of the total energy are two satellite peaks coming from the one photon
absorption and emission processes ( higher order processes are neglected
in our model as it has been discussed above).
As the bias increases, due to the asymmetry in the sample, the satellites   
become asymmetric too,
and for high bias only the satellite coming from the one photon
emission process shows up. 
In fig. 3.a we have plotted the coherent resonant tunneling current
density as a function of the external bias in the presence of the
electromagnetic field. The effect of the light on the current
density can be observed in fig. 3.b where 
the calculated current difference
between the case where  there is light present in the sample, minus
the case where there is not light present is drawn.
One observes a main structure in the region corresponding to the current
density threshold, the appearence of a shoulder for bias roughly at the 
center of the current peak and a smaller structure asociated with
the current cut off.\\
The change in the tunneling current as a function of the external bias
comes mainly from the change in the transmission coefficient where two
satellites appear corresponding to the one photon absorption and one
photon emission processes. The current is then obtained integrating to
all the available states with energies up to the Fermi level. In fig.
3.b a main peak shows up at an external bias smaller than the current threshold
bias for the case where no light is present in the sample. Physically it
comes from the fact that electrons in the emitter close to the Fermi
energy have a probability to absorb a photon and to tunnel through the
resonant state. Therefore the current increases in the presence of light
and the threshold bias for the current is smaller than the corresponding
to the case where the sample is not illuminated and a positive peak
appears in the current difference. For higher voltages, as the resonant
level crosses the Fermi energy, there is also an additional contribution to the
current coming from electrons absorbing a photon and tunneling non
resonantly through the double barrier. Finally, the physical reason for
the structure appearing at the current cut off bias (around .18 mV)
comes from the emission procceses once the resonant state in the well
crosses the bottom of the conduction band of the emitter.
This features are in good agreement with the experimental curve  \cite{3}.
\par
In order to compare with the experimental evidence we have to analyze
the sequential contribution to the tunneling current and confront it
with the coherent one. Therefore we have 
calculated the sequential tunneling current density in the presence of
light as well as the current difference with and without the photon
field ( fig. 4.a and 4.b respectively). One observes that the sequential
current falls down at the bias corresponding to the current cut off
more abruptly than the coherent one and that the current intensity is
of the same order as  the coherent one.  
More interesting is the fact that the current difference ( fig. 4.b )
for sequential tunneling is one order of magnitud smaller than
the corresponding to the coherent process (fig. 3.b) , therefore we conclude
that the experimental difference of currents, corresponds to the coherence 
tunneling proccess which dominates on the sequential one.
We have also evaluated the coherent and sequential current densities for the
same sample but considering photons with lower energy : $\hbar w=4.2
meV$ (fig. 5 and 6) in order to compare with the experimental results\cite{3} 
. In this case the same behaviour is observed as in the previous case
when the coherent contribution is compared with the sequential one : the
coherent tunneling current density is comparable in intensity with the sequential
one and the current difference (with and without light present in the
sample) is one order of magnitud larger in the coherent
proccess than in the sequential one, therefore  the last one is hidden
by the coherent contribution and is this one which should be compared
with the experiment. The agreement for this case (lower frecuency ) , is
not so good as for the previous one : the current difference for the
coherent case (fig. 5.b) presents a peak for a bias smaller than that
corresponding to the threshold current density whithout light . As in the
previous case, this structure comes from electrons close to the Fermi
level which absorb one
photon. This peak, which is  less intense than in the previous case
(less electrons with energies below $E_{F}$ than in the
case with higher photon energy) and narrower, is not observed
experimentaly , however the main features are well reproduced.
In order to see how it changes the relative intensity between coherent and
sequential tunneling current densities as a function of the barrier
thicknesses we have performed the same calculations as explained above for
thicker samples. For barrier thicknesses of $100 \AA$ and a  
well thickness of $50 \AA$ the same behaviour as before is observed, i.e.,
the coherent tunneling prevaleces 
in the current density difference with and without
light ( fig 7 and 8). We do not have considered the case for thinner 
barriers, because our model developed to describe the sequential 
tunneling cannot be applied properly to such cases and due to that 
we cannot make a correct comparison with the coherent tunneling case.
The reason is that we neglect in our model for sequential tunneling the
finite width of the resonant state in the well which increases as the barrier
thickness decreases.
\\
\par
Finally, we have analyzed the effect of an external magnetic field
applied in the current direction on this sample in the presence of the
photon field in the same configuration as in fig. 1.
We pay attention just to the coherent current which was dominant in
the absence of magnetic field.
We have analyzed two different cases: in the first one the
magnetic field intensity is $15.72 T$ and the photon energy $\hbar
w=13.6 meV$, therefore the cyclotron frecuency is twice the photon  frecuency. 
The second case corresponds to the same magnetic field intensity but a
photon energy of $\hbar w=27.2 meV$. For this magnetic field there are two
Landau levels which contributes to the current (fig 9.a). As the
electromagnetic field is schwitched on, the current density is modified
independently for each Landau level. This can be observed in fig. 9.b and
9.c. In the case that the cyclotron frecuency is twice the photon 
frequency (fig 9.b), the way the
light affects each Landau level separately is well resolved. In the
second case, where the cyclotron frequency is the same as the
 photon frequency (fig. 9.c) the current difference structures associated to
each Landau level overlap but remain decoupled each other. 
This result does not give any aditional
information to the photoassisted tunneling whithout magnetic field,  
because in this configuration the magnetic field only
affects to the planes parallel to the interfaces 
while the light affects the tunneling in the current direction. Therefore
the effect of both fields on the tunneling current are compleately
decoupled. More interesting would be to consider an electromagnetic field
with a component of the electric field in the interface planes. In this
case the effect on the current due to the magnetic field and the light
will not be independent each other anymore and new physical 
effects could be expected, that is the task of a next coming paper.
\section{CONCLUSIONS}
\par
We have studied the problem of coherent and sequential tunneling through a 
double barrier structure assisted by light considered to be present
All over the structure, i.e., emitter, well and collector, which is a
realistic description of the experiments.  By means of a
canonical transformation and time dependent perturbation theory up to first 
order, we have calculated the coherent transmission coefficient
and the electrical current through the system, for this specific problem,
resulting that the electromagnetic field couples states of different energies 
due to one photon absorption and emission processes. 
The higher order contributions to the current
(multiphoton absorption and emission processes) are much weaker and 
their contribution can be neglected in first aproximation.
As a result of that two satellite peaks appear in the transmission 
coefficient at both sides of the main resonant peak . Therefore, 
the total transmission coefficient and the coherent tunneling current are
affected by the photon field and new features in the current density are
observed. In order to obtain the total density current  
we have developed a model to analyze the sequential tunneling current
through a double barrier in the presence of light. We have calculated the
electronic tunneling current through the first barrier, i.e., from the emitter
to the resonant state in the well in the presence of light and the
current through the collector barrier coming from the electrons in the
well. The current conservation is reached when both currents are equal
and it determines the Fermi level in the well, i.e., the charge stored in
the well. The sequential contribution to the current coming out, 
is of the same order as the coherent one. For the current difference
with and without
electromagnetic field the coherent part is one order of magnitud larger
than the sequential one for the samples considered in our calculation.
Therefore it is the coherent current which should be compared with the
experiments \cite{3}, turning out to be both, theory and experiment , in good
agreement. We have also considered an external magnetic field applied in
the growth direction on
the double barrier structure and in the presence of light. The analysis
of the coherent tunneling current has been done for different ratios
between the cyclotron and the photon frecuencies.  \\ 
\newpage
\section{ACKNOWLEDGMENTS}
\par
One of us (Gloria Platero), acknowledges V.A. Chitta and T.Brandes for 
helpful discussions and the Theoretical Department in the 
PTB (Braunschweig) directed by
Prof. B.Kramer  for their hospitality and where part of this work has been 
done. This work has been supported in part by the Comision Interministerial
de Ciencia y Tecnologia of Spain under contract MAT 91 0210 and by
the Comission of the European Communities under contract SSC-CT 90 0201.\\

\newpage
\begin{figure}
\caption{Particle represented by a plane wave moving along the $z$ direction
crossing a DBS in the presence of an electromagnetic field polarized in the
$z$ direction}
\end{figure}

\begin{figure}
\caption{$Log_{10}$ of coherent  transmission coefficient as a function of 
total energy. ( $F=4.10^{5}V/m$, $\hbar w=13.6 meV$ ). a. 
Bias voltage $ V_{f}=0.0V$.  b. $V_{f}=0.10V$. c. $V_{f}=0.14V$.}
\end{figure}

\begin{figure}
\caption{a. Coherent tunneling current density as a function of voltage 
for light assisted tunneling. ($F=4. 10^{5} V/m$, $\hbar w=13.6 meV$). 
b. Current difference as a function of voltage between coherent light
assisted tunneling  and coherent tunneling without light present, 
( $F=4.10^{5}V/m$, $\hbar w=13.6 meV$).} 
\end{figure}

\begin{figure}
\caption{a. Sequential tunneling current density as a function of 
voltage for light assisted
tunneling. ($F=4. 10^{5} V/m$, $\hbar w=13.6 meV$). 
b. Current difference as a function of voltage between
sequential light
assisted tunneling  and sequential  tunneling without light present, 
( $F=4.10^{5}V/m$, $\hbar w=13.6 meV$) .} 
\end{figure}

\begin{figure}
\caption{a. Coherent tunneling current density as a function of 
voltage for light assisted
tunneling. ($F=4. 10^{4} V/m$, $\hbar w=4.2 meV$). 
b. Current difference as a function of voltage between coherent
 light assisted tunneling  and coherent tunneling without light present, 
( $F=4.10^{4}V/m$ and $\hbar w=4.2 meV$) .} 
\end{figure}

\begin{figure}
\caption{a. Sequential tunneling current density as a function of 
voltage for light assisted
tunneling. ($F=4. 10^{4} V/m$, $\hbar w=4.2 meV$). 
b. Current difference as a function of voltage between
sequential light
assisted tunneling  and sequential  tunneling without light present, 
( $F=5.10^{4}V/m$ and $\hbar w=4.2 meV$) .} 
\end{figure}

\begin{figure}
\caption{a. The same as fig. 3.a but for a barriers thickness of $100 \AA$.
b. The same as fig. 3.b but for a barriers thickness of $100 \AA$.}
\end{figure}

\begin{figure}
\caption{a. The same as fig. 4.a but for a barriers thickness of $100 \AA$.
b. The same as fig. 4.b but for a barriers thickness of $100 \AA$.}
\end{figure}

\begin{figure}
\caption{a. Coherent tunneling current density as a function of voltage 
for light assisted tunneling, in the presence of a magnetic field 
parallel to the current ($B=15.72T$, $\hbar w_{c}=27.2meV$, $F=4. 10^{5} V/m$, 
 $\hbar w=13.6 meV$). 
b. Current difference as a function of voltage between coherent light
assisted tunneling  and coherent tunneling without light present. For both
cases, in the presence of a parallel magnetic field.
 ($B=15.72T$, $\hbar w_{c}=27.2meV$, $F=4. 10^{5} V/m$, $\hbar w=13.6 meV$). 
c. Current difference as a function of voltage between coherent light
assisted tunneling  and coherent tunneling without light present. For both
cases in the presence of a parallel magnetic field.
 ($B=15.72T$, $\hbar w_{c}=27.2meV$, $F=4. 10^{5} V/m$, $\hbar w=27.2 meV$).} 
\end{figure}

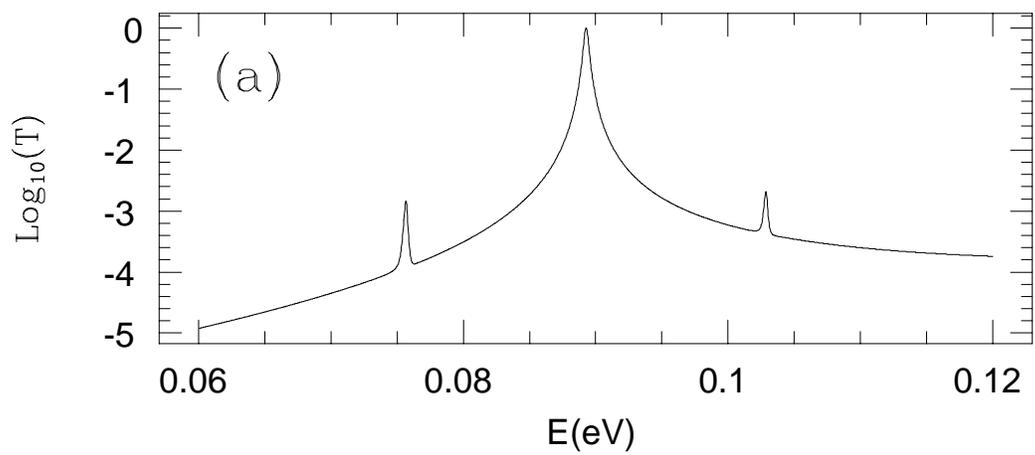
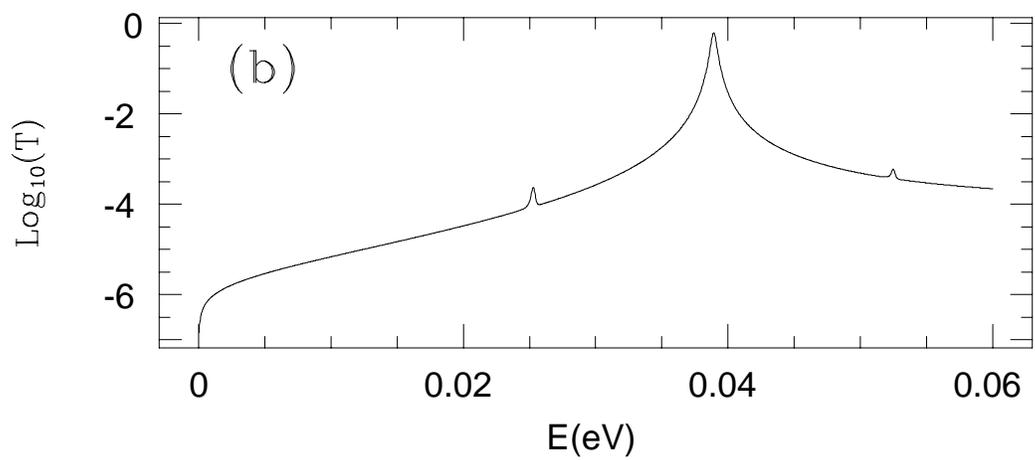
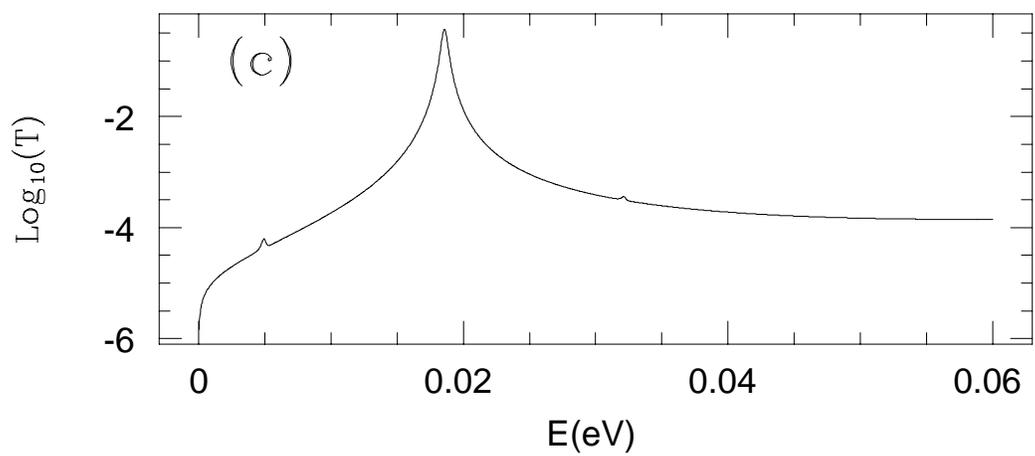
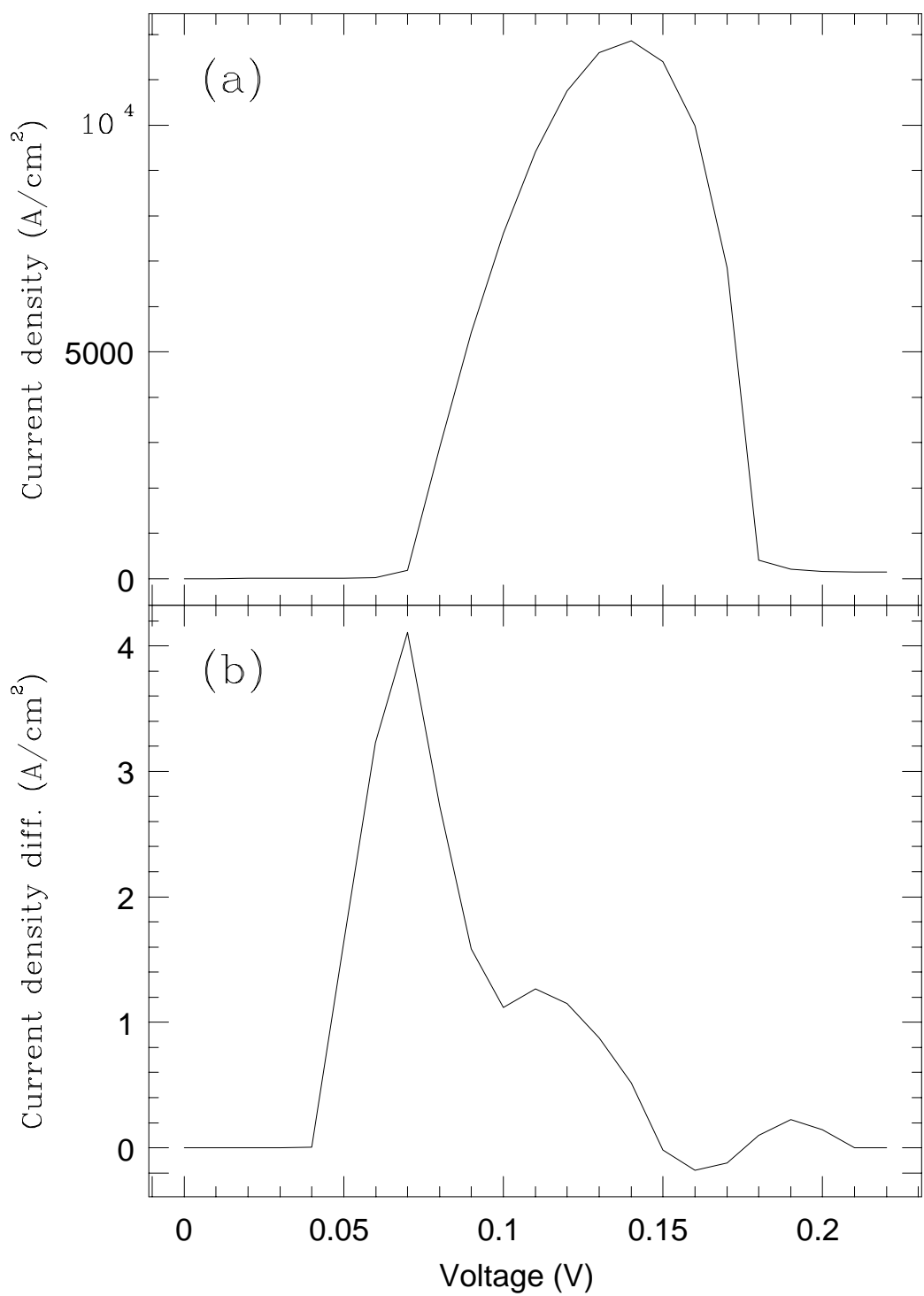
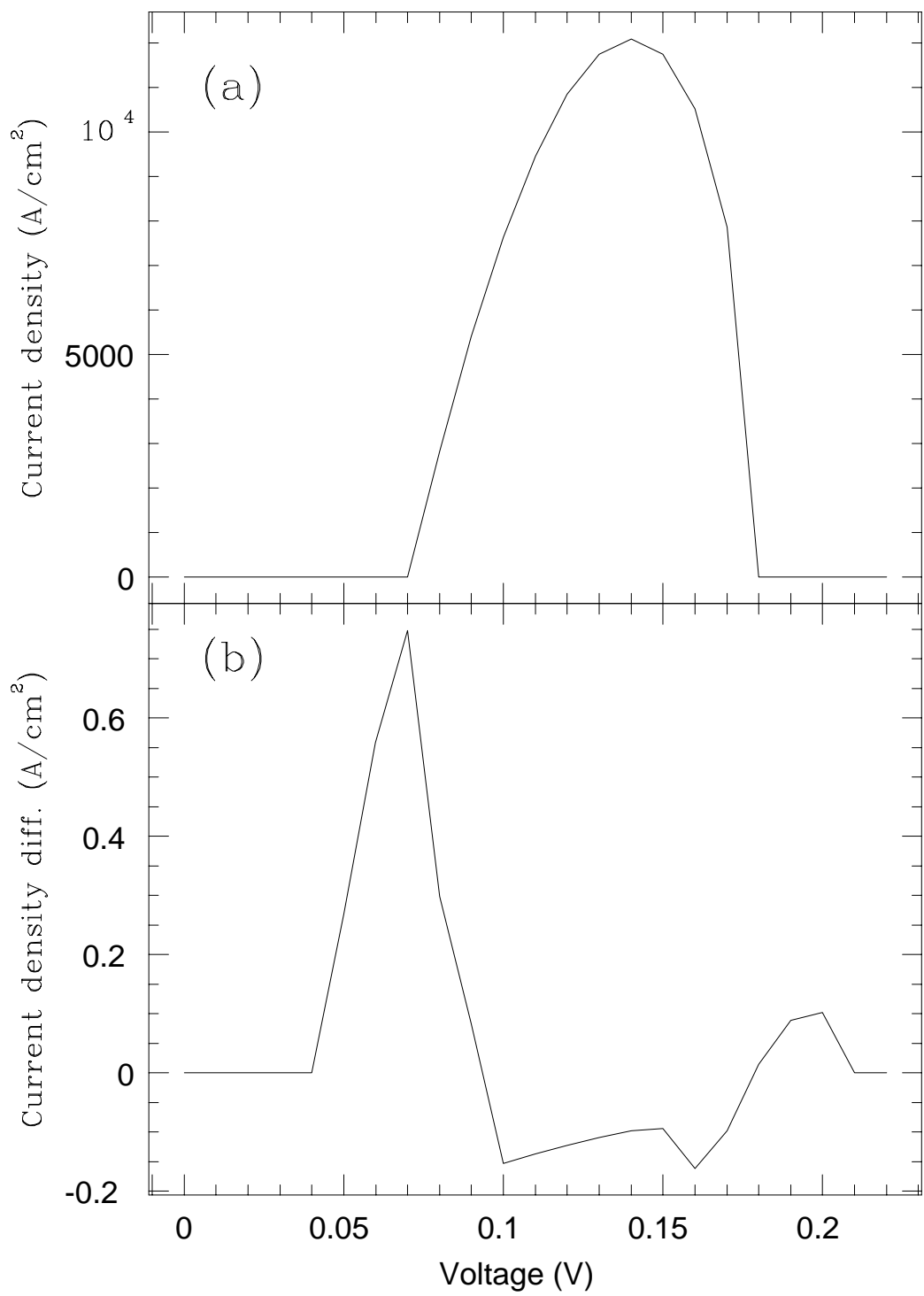
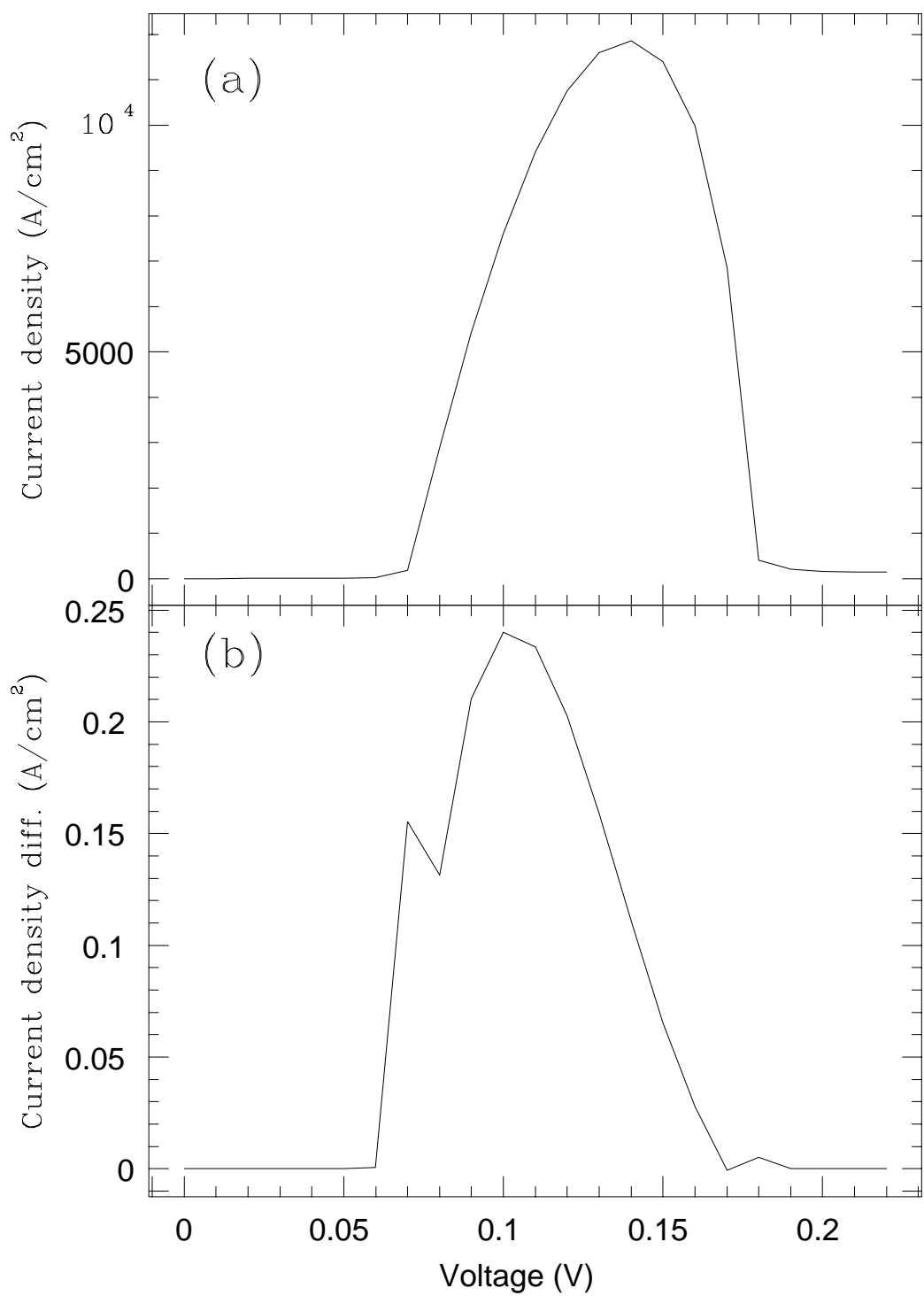
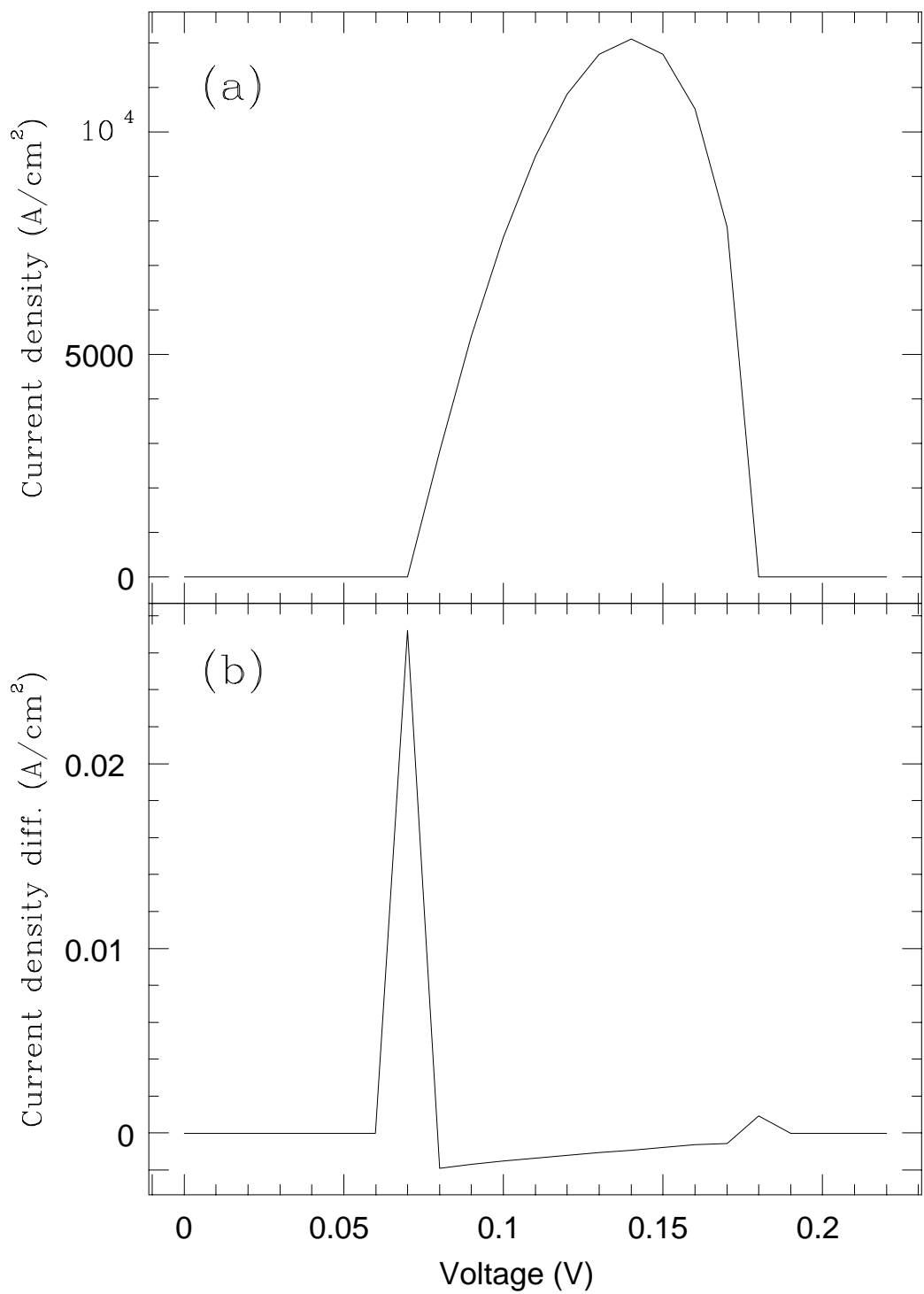
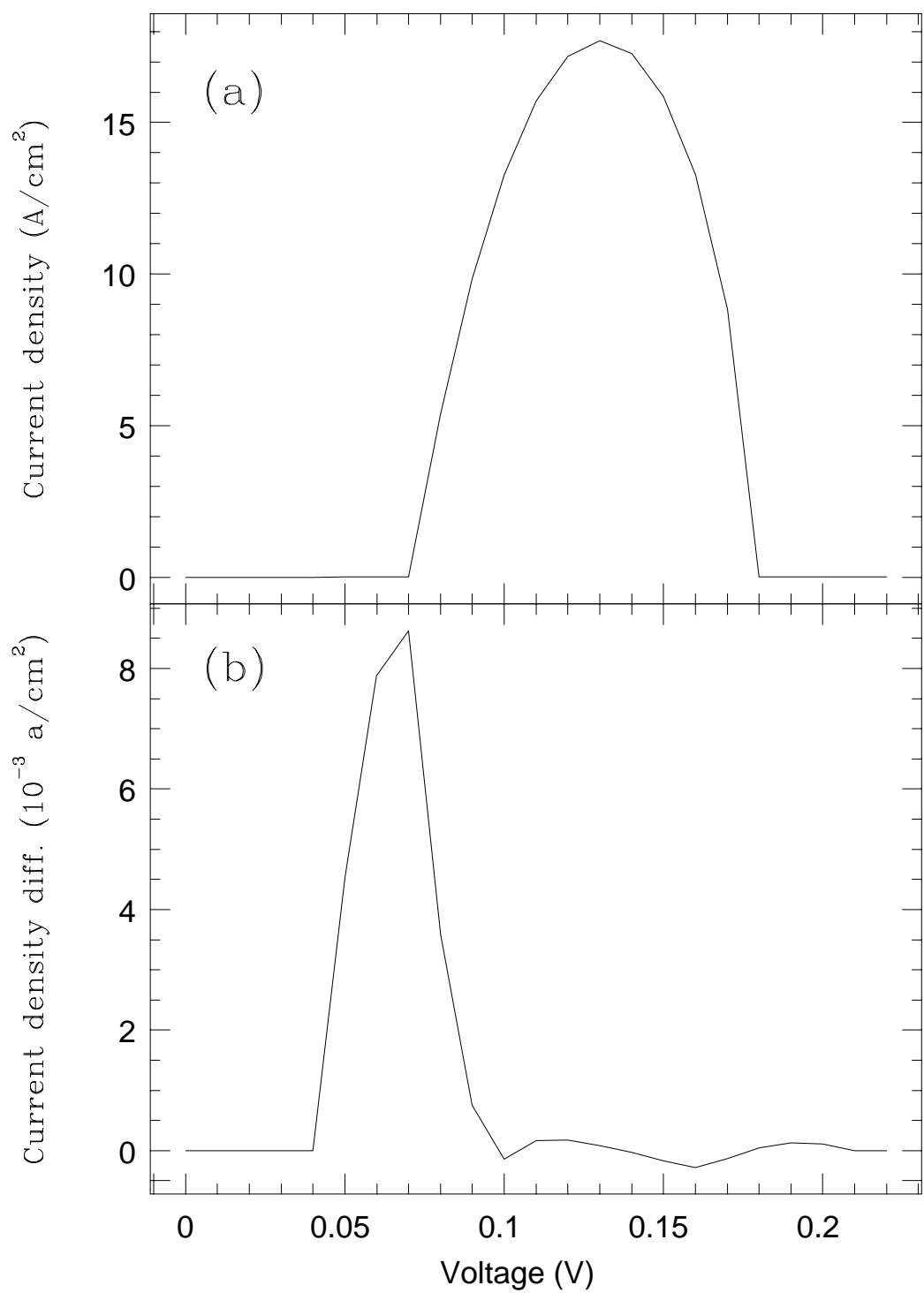
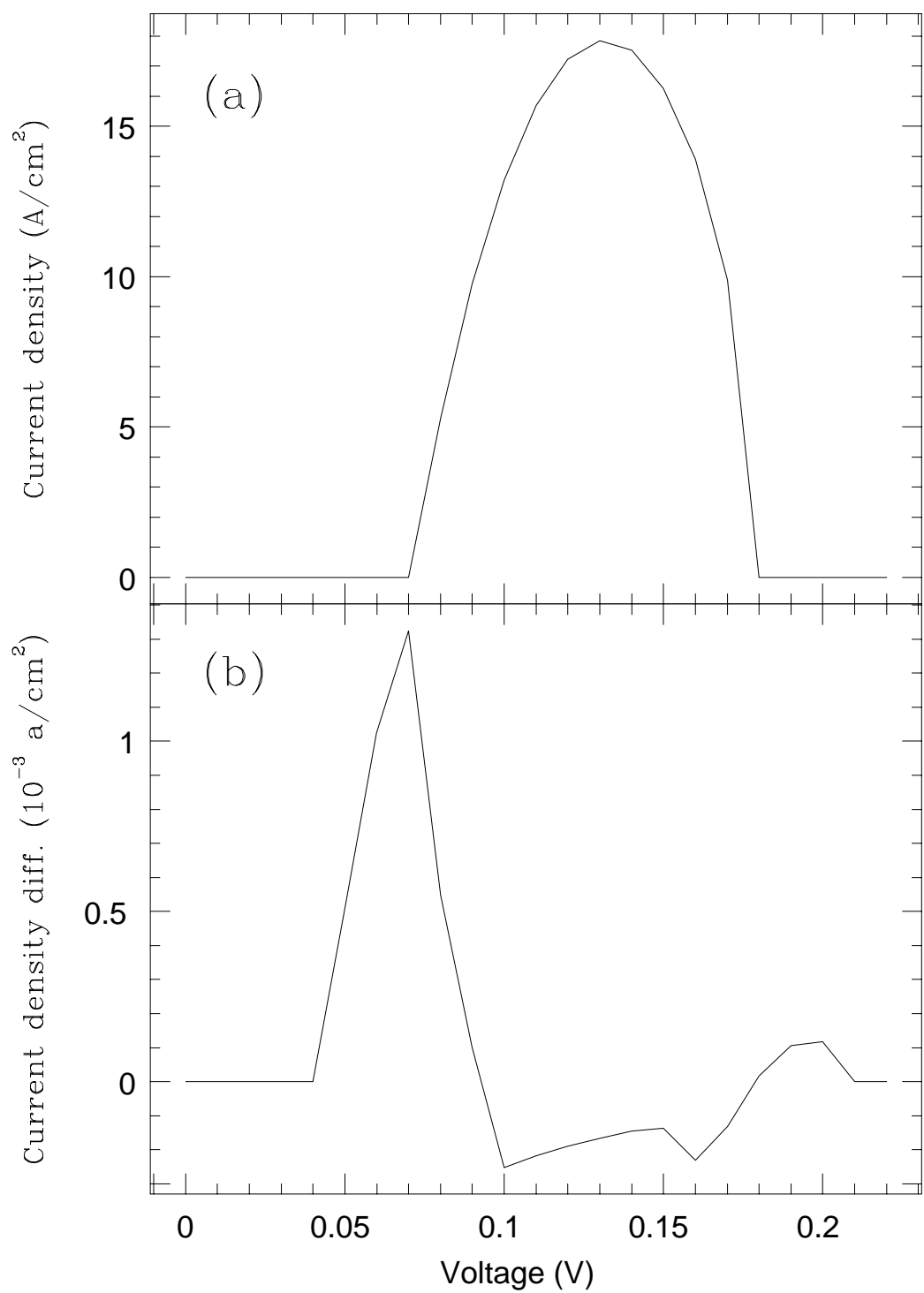
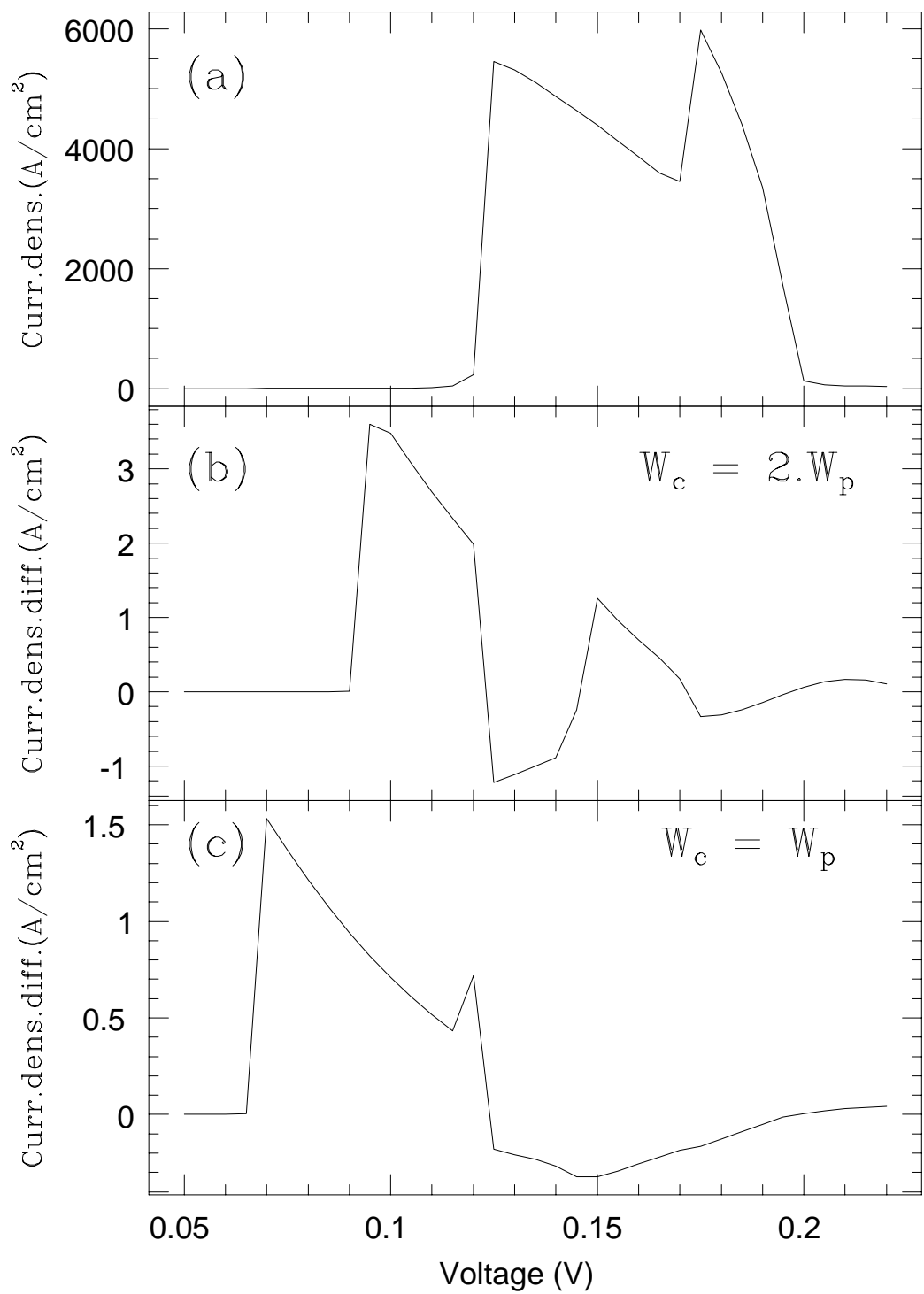
\end{document}